# A finite element method framework for modeling rotating machines with superconducting windings

R. Brambilla[1], F. Grilli[2], L. Martini[1], M.Bocchi[1], G.Angeli[1]

[1]Ricerca sul Sistema Energetico, Via Rubattino 54, 20134 Milano, Italy

[2]Karlsruhe Institute of Technology, Hermann-von-Helmholtz-Platz 1, 76344 Eggenstein-Leopoldshafen, Germany

## Abstract

Electrical machines employing superconductors are attractive solutions in a variety of application domains. Numerical models are powerful and necessary tools to optimize their design and predict their performance. The electromagnetic modeling of superconductors by finite-element method (FEM) is usually based on a power-law resistivity for their electrical behavior. The implementation of such constitutive law in conventional models of electrical machines is quite problematic: the magnetic vector potential directly gives the electric field and requires using a power-law depending on it. This power-law is a non-bounded function that can generate enormous uneven values in low electric field regions that can destroy the reliability of solutions.

The method proposed here consists in separating the model of an electrical machine in two parts, where the magnetic field is calculated with the most appropriate formulation: the *H*-formulation in the part containing the superconductors and the *A*-formulation in the part containing conventional conductors (and possibly permanent magnets). The main goal of this work is to determine and to correctly apply the continuity conditions on the boundary separating the two regions. Depending on the location of such boundary – in the fixed or rotating part of the machine – the conditions that one needs to apply are different. In addition, the application of those conditions requires the use of Lagrange multipliers satisfying the field transforms of the electromagnetic quantities in the two reference systems, the fixed and the rotating one. In this article, several exemplary cases for the possible configurations are presented. In order to emphasize and capture the essential point of this modeling strategy, the discussed examples are rather simple. Nevertheless, they constitute a solid starting point for modeling more complex and realistic devices.

## 1. Introduction

The worldwide production of high-temperature superconductor (HTS) tapes in the form of rare earth-based coated conductor has now exceeded several thousands of kilometers per year [1] and kilometer-long single pieces of tape with high performance and good uniformity are available from different manufacturers [2]. Thanks to the availability of the material, many HTS applications have been developed in recent years [3],[4]. Following this development, numerical models have become powerful and almost indispensable tools for designing reliable and efficient applications [5].
HTS windings are particularly attractive for HTS machines in virtue of the possibility of considerably reducing the applications' size and weight [6][7]. Generally, electrical machines consist of two main parts, a rotor and a stator, which rotate with respect to each other. In order to properly simulate the electromagnetic behavior of such machines (e.g. to calculate the dissipation caused by varying magnetic fields), finite-element-method models with rotating meshes are usually employed. At present, however, no such models exist for superconducting windings, and the simulation is typically performed with fixed geometry, see for example [8] . The reason lies in the very non-linear behavior of the resistivity of the superconductors, which makes the use of conventional models difficult or impossible.

With this paper, we introduce such possibility. The main idea of this work is to divide the geometry of the machine in two different parts. A first part, containing the conventional conductors (if any), where the electromagnetic field is computed by means of the magnetic vector potential formulation (*A*-formulation). A second part, containing the superconductors, where the electromagnetic field is directly obtained by solving Maxwell's equations (*H*-formulation). The *H*-formulation is a widely-used framework that resulted to be appropriate for the numerical application of the power-law that is normally used in characterizing the resistivity of superconductors.

By this method a fully HTS machine involves a common boundary line that separates the *H*-formulation regions from the *A*-formulation regions. On this line the coupling conditions of these two formulations have to be imposed and their formulation is the central object of this study.

**POWER-LAW**. In order to simulate superconductors by finite element method, it is necessary to model their electric behavior with a non-linear conductivity/resistivity, often in the form of a power-law [9], which can be expressed in the following algebraic equivalent forms

(i) $\quad \rho(J) = \rho_0 |J/J_c|^{n-1}$ $\qquad$ (ii) $\quad \sigma(J) = \sigma_0 |J/J_c|^{1-n}$

(iii) $\quad \rho(E) = \rho_0 |E/E_c|^{1-1/n}$ $\qquad$ (iv) $\quad \sigma(E) = \sigma_0 |E/E_c|^{1/n-1}$

with $\rho_0 = 1/\sigma_0 = E_c/J_c$, where $J_c$ is the critical current density, $E_c$ is an arbitrary electric field criterion (usually set equal to $10^{-4}$ V m$^{-1}$), and $n$ is the power index defining the steepness of the experimental current-voltage curve used to characterize the superconductors. Despite the equivalence of (i-iv), in FEM applications the choice of the independent variable to be used in the power-laws (*J* or *E*) is imposed by the necessity of avoiding circular definitions.

With a vector potential formulation, the electric field $\boldsymbol{E}$ is a primary quantity since it is the time derivative of the vector potential and does not require any constitutive relation that may create circularity. Hence we can write

(1) $\quad \sigma(\text{E}) \, \partial_t \boldsymbol{A} - \frac{1}{\mu} \nabla^2 \boldsymbol{A} = \boldsymbol{J_e}$

(2) $\quad \boldsymbol{B} = \nabla \times \boldsymbol{A}, \quad E = -\partial_t \boldsymbol{A}, \quad J = \sigma(E) \boldsymbol{E} + \boldsymbol{J_e}$

Therefore (iv) can be used without circularity.

On the contrary, in a formulation based on the magnetic field components (using directly Maxwell's equations without any potential, the so-called *H*-formulation), the current density $\boldsymbol{J}$ is a primary quantity, because it is a combination of the spatial derivatives of the magnetic field components without any additional constitutive relation. Hence

(3) $\quad \mu \, \partial_t \boldsymbol{H} + \nabla \times \boldsymbol{E} = 0$

(4) $\quad \boldsymbol{J} = \nabla \times \boldsymbol{H}, \quad \boldsymbol{E} = \rho(J) \boldsymbol{J}$

As a consequence, the formula (i) can be used without circularity.

From the graphs of (i) and (iv) in Figure 1 one can immediately observe that $\rho(J)$ is a low increasing function near the origin (J=0) which is usually the condition at the beginning of modeling; on the contrary, $\sigma(E)$ is not bounded in the origin and is very steep. This different behaviour is the most relevant issue in numerical applications. Also, as noted by Rhyner [9], the power form of (iv) is much more delicate to be treated numerically. The robust form of (i) on the contrary does not require any peculiar handling.

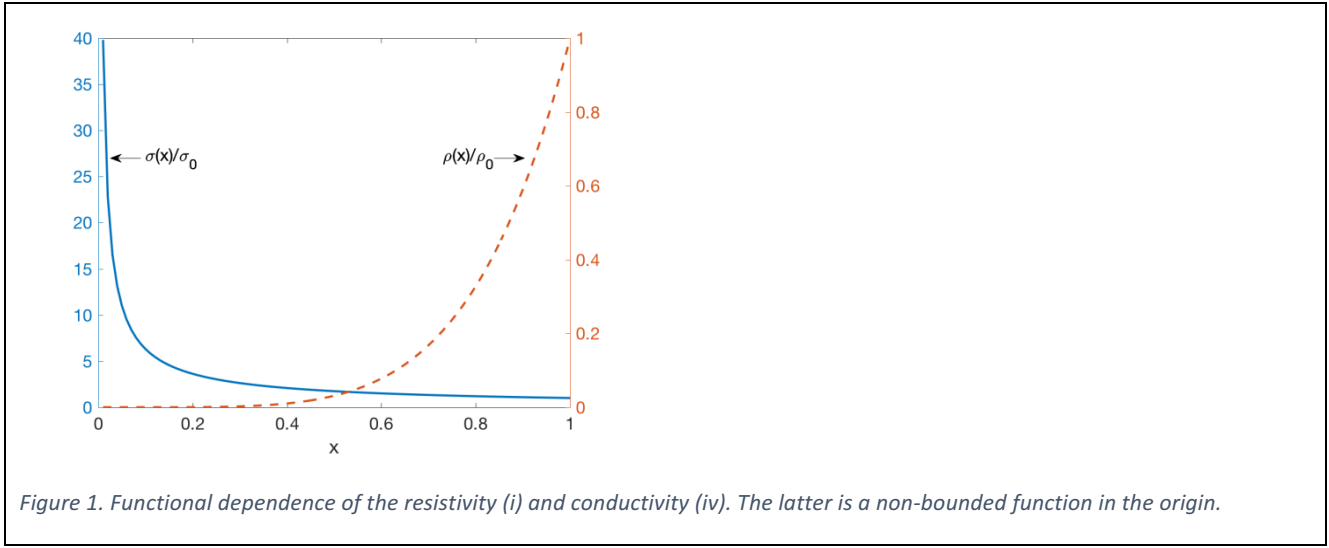

*Figure 1. Functional dependence of the resistivity (i) and conductivity (iv). The latter is a non-bounded function in the origin.*

For our purpose in the following pages we need to make reference to two essential boundaries.

**TRAMODEL-LINE**. A complex system containing both normal and superconducting conductors can be conveniently divided in different parts (domains), each containing conductors of one type only. As a consequence, the magnetic field in the domains with normal conductors can be computed by the usual magnetic vector potential *A* and for brevity we call these domains "*A*-potential parts". In the domains containing the superconductors, the magnetic field can be directly computed using Maxwell's equations (without any potential) and we call these domain "*H*-components parts". We shall denote the common boundary(-ies) between these different type of parts as the _tramodel-line(s)_. The major difficulty consists in finding a way to properly couple the electromagnetic quantities across that line, so that the requested continuity of electromagnetic conditions is correctly satisfied. With FEM models, simply imposing the equality of the electromagnetic quantities on the common boundary of the two formulations resulted to be inadequate to force their continuity. In order to achieve this request, it was necessary to reformulate the boundary conditions in equivalent source terms to be expressed in weak form.

**ROTATION-LINE.** In the case of rotating machines the rotating part and the fixed part are separated by a circular line that we shall call _rotation-line._ In FEM this circle separates the fixed mesh from the rotating mesh and other appropriate coupling condition have to be imposed to maintain the continuity of the field quantities formulated in the fixed and the rotating coordinate system. The simplest solution is to divide the geometry so that the rotation-line is wholly inside the *A*-potential parts since in this case it results to be a scalar continuity line of the *A*-potential in the two coordinate systems. In 2D models (to which we refer in this paper) the scalar continuity refers only to the $A_z$ component

$$A_z(x,y) = A'_z(x',y')$$

Depending on the different possible arrangement of normal/superconducting and fixed/rotating parts, we will have four possible configurations with as many conditions of coupling. In the second part of this work we shall give schematic examples of all these possibilities.

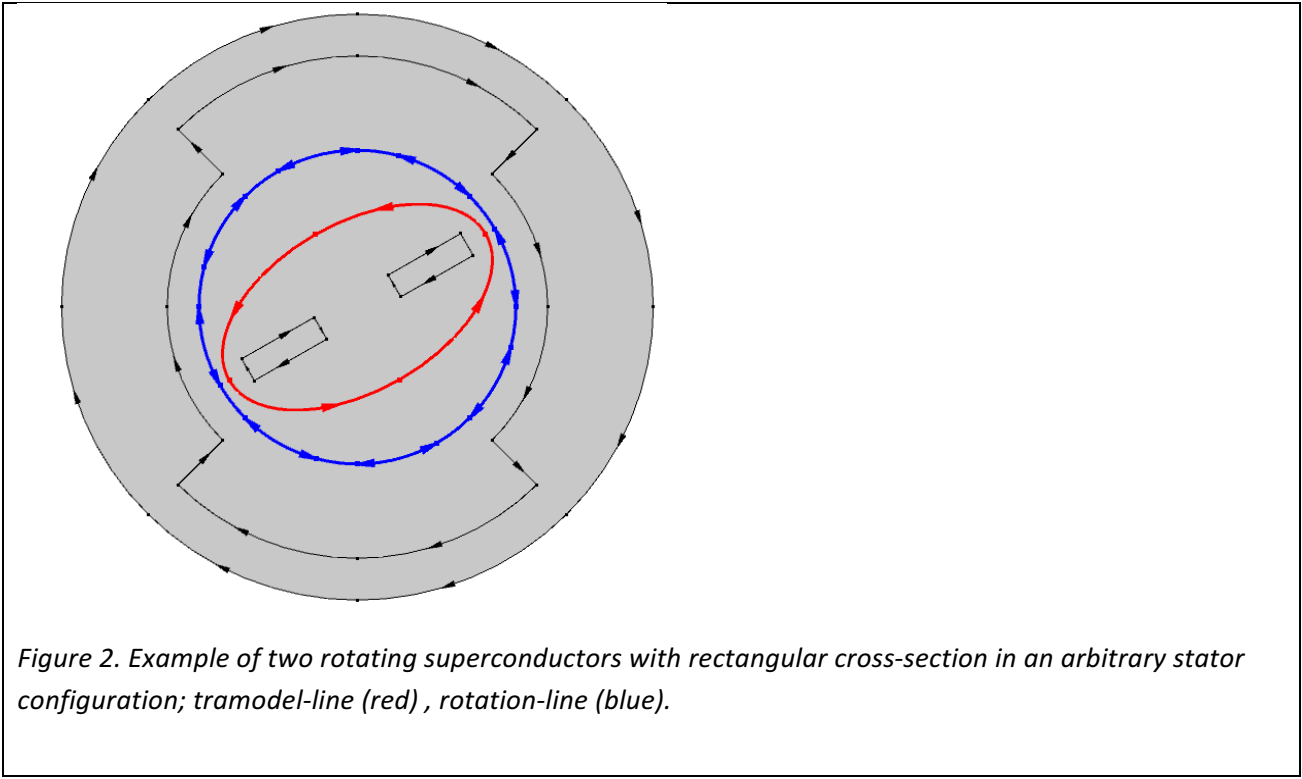

*Figure 2. Example of two rotating superconductors with rectangular cross-section in an arbitrary stator configuration; tramodel-line (red) , rotation-line (blue).*

As a schematic example, in the simple model shown in Figure 2, the domains inside the red ellipse (containing two superconductors with rectangular cross section and a part of air trafer gap) are the *H*-component parts. The outer domains (remaining trafer gap and iron stator poles) are the *A*-potential parts. Note that the blue circular rotation-line, separating the external fixed part from the internal rotating part, is entirely placed in the potential part. On the contrary the tramodel-line, i.e. the red ellipse separating the *H*- and *A*-formulations, where the coupling conditions have to be imposed is wholly in the rotating part.

The main problem with mixed formulations for conductive (superconducting) and non-conductive domains is the coupling of fields on the common boundaries. See for instance the extensive analysis by Birò [14], who consider the coupling between different potentials formulations. We here consider a simpler case where only one potential exists (for non-conductive/conductive regions) and direct components for superconducting regions.

## 2. Coupling the H- and A- formulations

In order to frame the analysis of formulations coupling, let us start considering a simple 2D case where the *xy* plane is divided in two regions $\Omega$ and $\Omega'$ by a closed boundary $\Gamma$ (the tramodel-line), which we will further assume to be arbitrarily divided in the two parts $\Gamma_B(t)$ and $\Gamma_H(s)$ (*Figure 3*), where *t* and *s* are appropriate parameters defining the points on these boundaries. Our goal is to describe the regions $\Omega$ and

$\Omega'$ with the $A$- and $H$-formulations, respectively. The space is considered here as a region of high electrical resistivity **Error! Reference source not found.**. The models considered below are 2D cross sections, with in-plane magnetic field and out-of-plane currents.

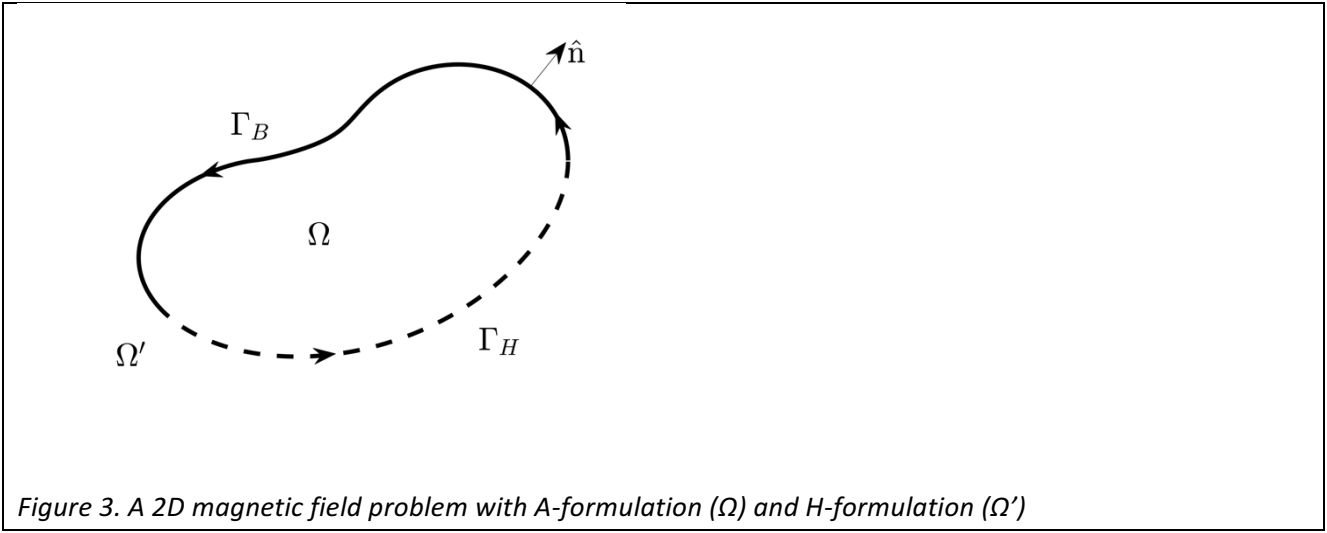

*Figure 3. A 2D magnetic field problem with A-formulation ($\Omega$) and H-formulation ($\Omega'$)*

**Region $\Omega$ ($A$-formulation).** The current density $\boldsymbol{J}$ has only the component $J_z(x, y)$ as does the electric field $E_z(x, y)$ and, therefore, the magnetic vector potential only needs $A_z(x, y)$. The electromagnetic quantities are given by

(5)     $B_x = \partial_y A_z, \quad B_y = -\partial_x A_z, \quad E_z = \partial_t A_z, \quad J_z = \sigma E_z$

Let us now consider the possible boundary conditions.

On the boundary part $\Gamma_H(s)$ we choose to impose the tangential component of the magnetic field to be a given function: $H_t(s) = f(s)$. Since $H_t = (\boldsymbol{t} \cdot \boldsymbol{H}) = (\boldsymbol{H} \times \boldsymbol{n})_z = (0,0, H_x n_y - H_y n_x)$ using (5) we have

(6)     $\mathbf{n} \cdot \nabla A_z(s) = \partial_n A_z(s) = \mu f(s) \qquad$ (on $\Gamma_H$)

which is clearly a Neumann boundary condition for $A_z$. Hence imposing a tangential magnetic field corresponds to a Neumann boundary condition.

On the other part $\Gamma_B(t)$ we can impose a Dirichlet boundary condition. As a matter of fact, the normal component of the magnetic induction $\boldsymbol{B}$ can be freely assigned $\boldsymbol{B} \cdot \boldsymbol{n} = b(t)$, i.e.

(7)     $B_n(t) = (\nabla \times \boldsymbol{A}) \cdot \boldsymbol{n} = \nabla \cdot (\boldsymbol{A} \times \boldsymbol{n}) = b \ (t) \qquad$ (on $\Gamma_B$)

In other words, imposing $\boldsymbol{A} \times \boldsymbol{n} = \boldsymbol{\alpha}\,(t)$   we have $\nabla \cdot \boldsymbol{\alpha}\,(t) = b(t)$. This means that, in particular cases, given *b(t)*, one can derive $\boldsymbol{\alpha}(t)$ and from it the boundary conditions for $\boldsymbol{A}$. Therefore, in a 2D case, equation (7) reduces to the Dirichlet boundary condition $A_z(t) = \alpha(t)$.

The simplest case $b(t) = 0$ represents the boundary condition $B_n(t) = 0$, i.e. the condition of *magnetic insulation* along the boundary $\Gamma_B$. In this case by (7) we have $\nabla \cdot (\boldsymbol{n} \times \boldsymbol{A}) = 0$, which can be satisfied by $A_z(t) = $ constant  along $\Gamma_B(t)$. Therefore, along this boundary part there is only the tangential component of the magnetic field $\boldsymbol{B_t} = \left(B_x t_x, B_y t_y, 0\right)$.

Another simple case that can be easily brought to a Dirichlet boundary condition is that where the boundary part $\Gamma_B(t)$ is exposed to an <u>external uniform magnetic field</u> $\boldsymbol{B_0} = \left(B_{0x}, B_{0y}\right)$. In this case the condition (7) is equivalent to the Dirichlet boundary condition $A_z = yB_{0x} - xB_{0y}$ along $\Gamma_B(t)$. In general, in the region $\Omega$ we will have a mixed boundary value problem for the potential $A_z$

$$\text{(8)} \quad \begin{cases} A_z = b(t) & \text{(on } \Gamma_B(t)) \\ \partial_n A_z = \mu f(s) & \text{(on } \Gamma_H(s)) \end{cases}$$

It is also possible, as displayed in Figure 4, to bend the boundary part $\Gamma_H(s)$ in a way to obtain a closed loop containing the region $\Omega'$ so that the boundary part $\Gamma_B(t)$ becomes the external boundary. This is the case of rotating machines.

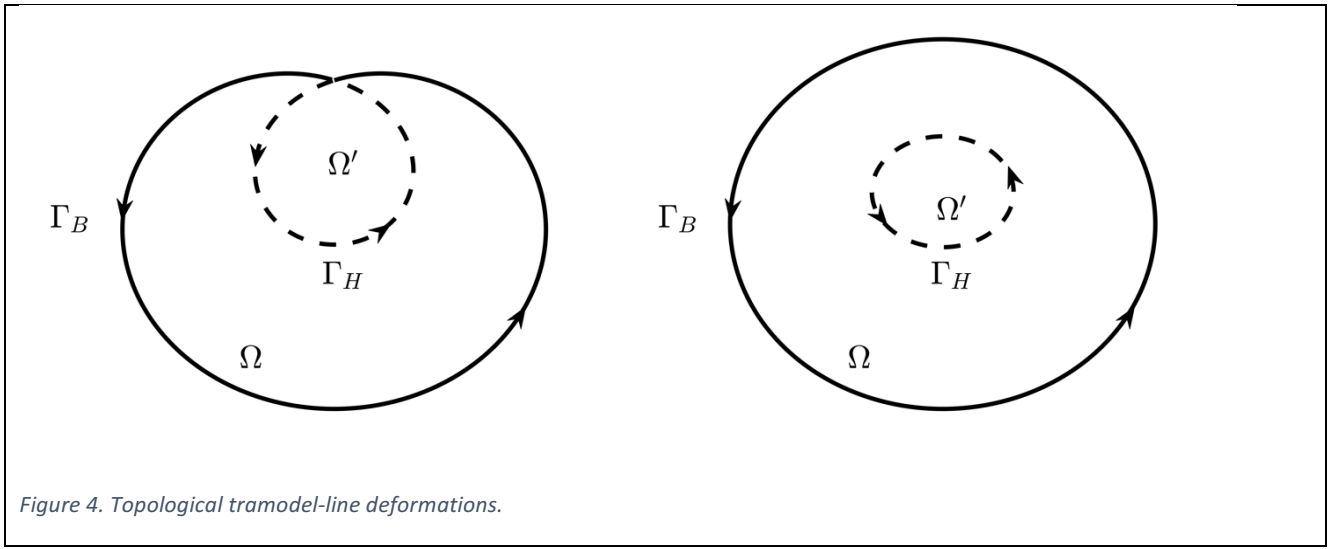

Figure 4. Topological tramodel-line deformations.

**<u>Region $\Omega'$ ($H$-formulation)</u>**. Since the magnetic field is solved with the *H*-formulation on both parts of the boundary, one has to give the value of the tangential field, i.e. $\boldsymbol{H_t}(\Omega') = \boldsymbol{H_t}(\Omega)$. Therefore

$$\text{(9)} \quad \begin{cases} H_t = H_x t_x + H_y t_y = -\sigma \partial_t A_z & \text{(on } \Gamma_B(t)) \\ H_t = H_x t_x + H_y t_y = f(s) & \text{(on } \Gamma_H(s)) \end{cases}$$

where $f(s)$ is the same current density used in the region $\Omega$. It can be a known function (input of the problem) or the result of calculation if there are field sources. If the boundary $\Gamma_B$ represents a condition of magnetic insulation, we will simply have

$$\text{(10)} \quad \begin{cases} H_x n_x + H_y n_y = 0 & \text{(on } \Gamma_B(t)) \\ H_x t_x + H_y t_y = f(s) & \text{(on } \Gamma_H(s)) \end{cases}$$

The equations (3) of the *H*-formulation are therefore coupled both by the presence of the electric field inside $\Omega$ and by the boundary values imposed by the tangential components. Nevertheless, the problem is of Dirichlet type on the whole boundary.

We shall now apply these general considerations to the simple 2D model of Figure 5(a), the case of a rectangular superconducting tape carrying a time-dependent transport current $I_a(t)$ and a round metallic conductor carrying a current $-I_a(t)$. We divide the geometry in two parts, drawing a curve around the superconducting tape (for example an ellipse of arbitrary dimensions), which will serve as tramodel-line. We calculate the field in the domain outside this line with the $A$-potential formulation approximated in FEM with lagrangian elements, whereas for the two domains (superconductor and air) inside it we will use the $H$-components formulation approximate in FEM by edge elements.

The total current flowing in the superconductor (which is in the part with the $H$-formulation) can be imposed by an integral constraint [10]

(a)

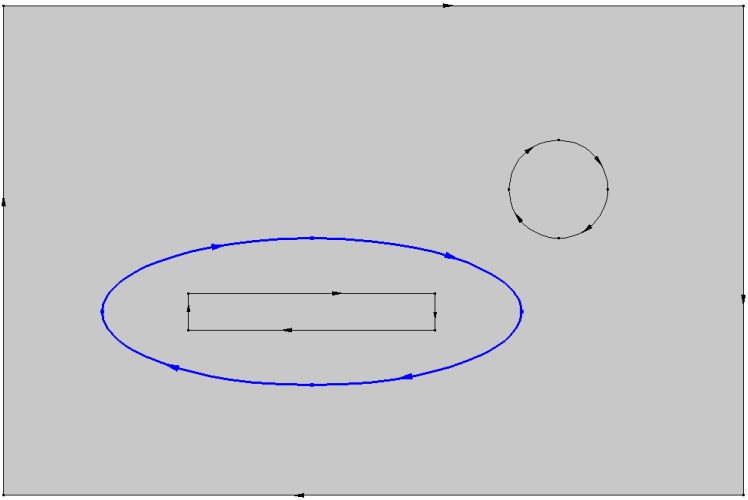

(b)

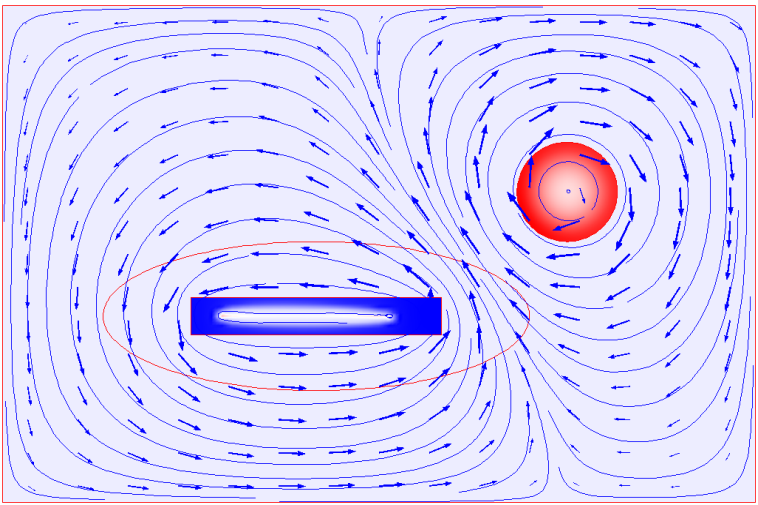

(c)

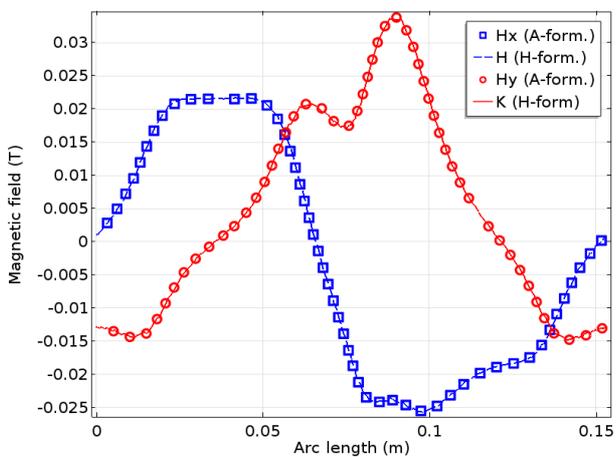

*Figure 5. (a) Definition of the tramodel-line (blue ellipse) in the case of a rectangular superconductor and a round metallic wire. (b) Magnetic field lines and current density distributions. (c) Magnetic field components along the tramodel-line, calculated with the A- and H-formulation: the overlapping of the curves validates the continuity of the magnetic field across the tramodel-line.*

(11)     $I_{SC} = \int_{SC} J_z(t)\mathrm{d}s = I_a(t)$

The return current in the metallic wire (which is in the part with the *A*-formulation) is imposed by means of an aptly sized *uniform* external current density $J_e(t)$ in its cross-section

(12)     $J_e(t) = \sigma\, V_{\mathrm{wire}}(t)/L$

where $L$ is a longitudinal reference length (usually set equal to 1 m), $\sigma$ is the electrical conductivity of the metal and $V_{\mathrm{wire}}$ is the voltage drop (across $L$) necessary to impose the return current. This is obtained by adding the global equation

(13)     $I_{\mathrm{wire}}(t) = -I_a(t) = -\sigma \int_{\mathrm{wire}} \partial_t A_z(t)\mathrm{d}s - \mathrm{sez} * J_e(t)$
.

Since we consider a magnetically insulated system, we can impose $A_z = 0$ on the outer boundary. In order to couple the two formulations, on the tramodel-line it is necessary to define the tangential component of the magnetic field in the two formulation. In the *A*-formulation we define

$H_t^{(A)} = t_x \cdot H_x + t_y \cdot H_y$

and in the *H*-formulation

(14)     $H_t^{(H)} = t_x \cdot H + t_y \cdot K$

(for convenience, we call $H = H_x$ and $K = H_y$ the magnetic field components in the *H*-formulation part to distinguish them from the corresponding $H_x$ and $H_y$ in the *A*-formulation part).

In our experience of FEM modeling (using the Comsol Multiphysics [11] package), the naïve coupling $H_t^{(A)} = H_t^{(H)}$ on ellipse turns out to be ineffective to link the two fields and a stronger condition to pass values between regions to be connected has to be devised.

On such boundary, we have found that the coupling conditions is effective if we impose (in weak form) in the two formulations

(15)     *A*-formulation:          $\mathrm{H}_t^{(H)} \cdot \mathrm{test}(\mathrm{A}_z)$

(16)     *H*-formulation:          $: \mathrm{E}_z \cdot \mathrm{test}\left(\mathrm{H}_t^{(H)}\right)$

Their meaning is to force the common values as sources: $H_t^{(H)}$ acts as a source for *A*-formulation part and $-\partial_t A_z = E_z$ as a source for the *H*-formulation part.

The resistivity $\rho(J)$ of the superconductor is given by the power-law (i) (which is possible to use because the superconductor lies within the *H*-formulation part). The resistivity of the normal cylinder is a constant. The continuity between the two formulations is excellent: the field lines exhibit no sign of distortion across the red tramodel-line. This is shown in detail in Figure 5(c), which plots the two field components $(H_x, H_y)$ along the tramodel-line, calculated with the two formulations. Their coincidence confirms the effectiveness and correctness of the devised coupling conditions.

With the same general settings, it is possible to change the potential and the component parts. The only essential modification is the change of sign in the joining conditions (15) and (16), which is due to the change of direction of the tramodel-line as seen by the two formulations.

# 3. Application to rotating systems

In order to model a rotating machine, it is convenient to use two reference systems, one for the fixed part (spatial frame) and one for the rotating part (material frame). We will use the coordinates of spatial type $(x, y)$ and the coordinates of material type $(X, Y)$. In the case of a planar rotation, the temporal coordinates are linked through the angle $\omega t$.

$$(17) \qquad \begin{pmatrix} X(x, y, t) \\ Y(x, y, t) \end{pmatrix} = T \begin{pmatrix} x \\ y \end{pmatrix} \qquad \text{where} \qquad T = \begin{pmatrix} \cos(\omega t) & \sin(\omega t) \\ -\sin(\omega t) & \cos(\omega t) \end{pmatrix} \qquad .$$

The vectors undergo a similar transformation. In particular, the magnetic field

$$(18) \qquad \begin{pmatrix} B_X \\ B_Y \end{pmatrix} = T \begin{pmatrix} B_x \\ B_y \end{pmatrix}.$$

The field equations are solved in both reference systems, and therefore the main problem is the transformation of the physical quantities from one reference system to the other one. In particular, in the case of non-relativistic speeds, the Lorentz transforms for the electric and magnetic fields reduce to the following simple relations

$$\boldsymbol{B}_{\text{spatial}}(x, y) = \boldsymbol{B}_{\text{material}}(X, Y)$$

$$\boldsymbol{E}_{\text{spatial}}(x, y) = \boldsymbol{E}_{\text{material}}(X, Y) - \boldsymbol{v}(x, y) \times \boldsymbol{B}_{\text{material}}(X, Y)$$

where $\boldsymbol{v}(x, y)$ is the velocity of a point $(x, y)$ of the rotor (in the spatial frame), i.e. $\boldsymbol{v} = (v_x, v_x, 0) = \boldsymbol{\omega} \times \boldsymbol{x} = (-\omega y, \omega x, 0)$. In the 2D case we will have (using a simplified notation)

$$(19) \qquad \begin{pmatrix} B_x \\ B_y \end{pmatrix} = T^{-1} \begin{pmatrix} B_X \\ B_Y \end{pmatrix}$$

$$(20) \qquad E_z = E_Z + \omega(X B_X + Y B_Y) = E_Z + \mathrm{d}E_Z.$$

In passing from one reference system to the other one, the magnetic field simply rotates, and its magnitude does not change. On the contrary, the electric field varies by $\mathrm{d}E_Z$, due to the rotation. Inverting equation (20) we have

$$(21) \qquad E_Z = E_z - dE_z = E_z - \omega(x B_x + y B_y)$$

The relations above will have to be applied only if the tramodel-line is in the rotating part of the model. In addition, one can show that along an arbitrary curve the tangent vector transforms like $\boldsymbol{t}_m = T \boldsymbol{t}_s$ and therefore the tangential field $H t_m = \boldsymbol{t}_m \cdot \begin{pmatrix} B_X \\ B_Y \end{pmatrix} = \boldsymbol{t}_s \cdot \begin{pmatrix} B_x \\ B_y \end{pmatrix} = H t_s$ is identical in the two reference systems.

## 3.1    Modelling a rotating machine

A rotating machine consists of two essential parts: the stator and the rotor, which are separated by an air gap. In order to apply the finite element method to model the machine, it is necessary to handle separately the stator and the rotor, by constructing two separated meshes, so that the rotor's one can rotate [11].

An example is shown in Figure 6.

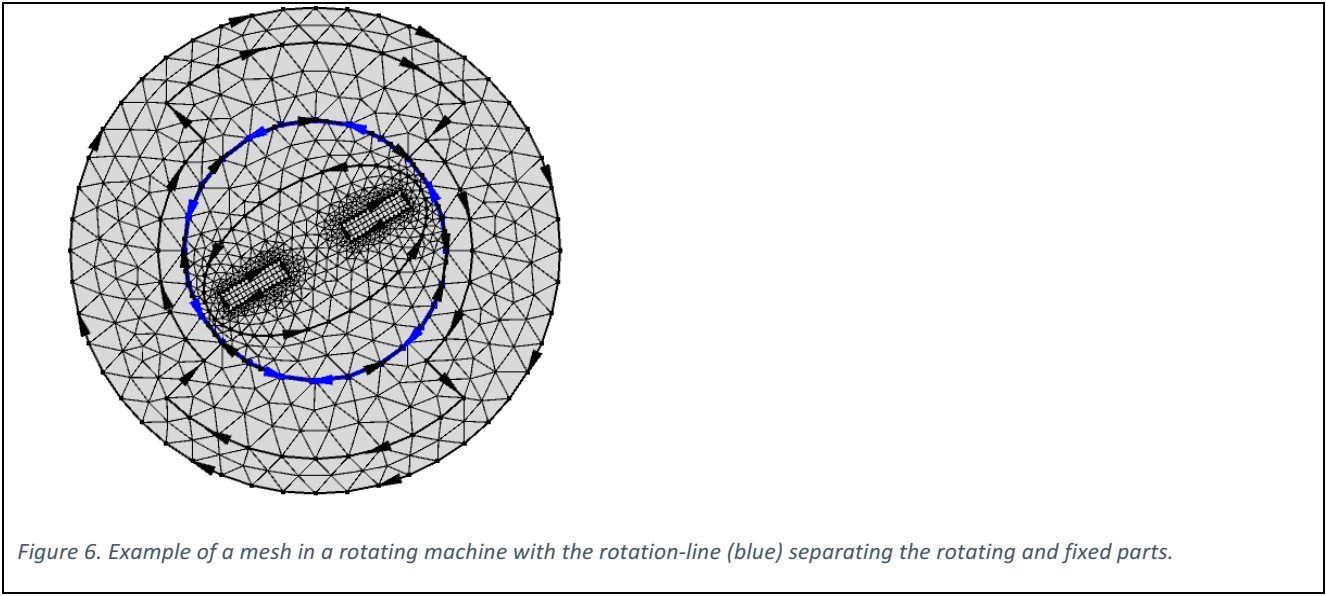

*Figure 6. Example of a mesh in a rotating machine with the rotation-line (blue) separating the rotating and fixed parts.*

The continuity of physical variables is obtained in the simplest way if the rotation-line is entirely inside domains with the same formulation. In 2D problems we choose the $A$-formulation, since $A_z$ is a scalar type variable so it does not require direction transform as $H$-formulation requires. Then there will be two possibilities:

   *a*) the tramodel-line occurs in the rotating part

   *b*) the tramodel-line occurs in the fixed part.

The joining condition are different in the two cases, because they will need to be explicitly expressed in the rotating system (case *a*) or in the fixed system (case *b*).

In the following, we will analyze these different possible configurations in simple examples that can be used as reference templates for more realistic models.

## 3.2 Model 1 (rotating tramodel-line)

Figure 7 shows the cross-section of a machine, where the rotor is composed by a superconducting winding represented by the rectangular cross sections with an iron nucleus and the stator is an iron tube with four polar expansions. The geometry is built such that the rotation-line separating the fixed and rotating parts passes through the polar expansions (blue line).

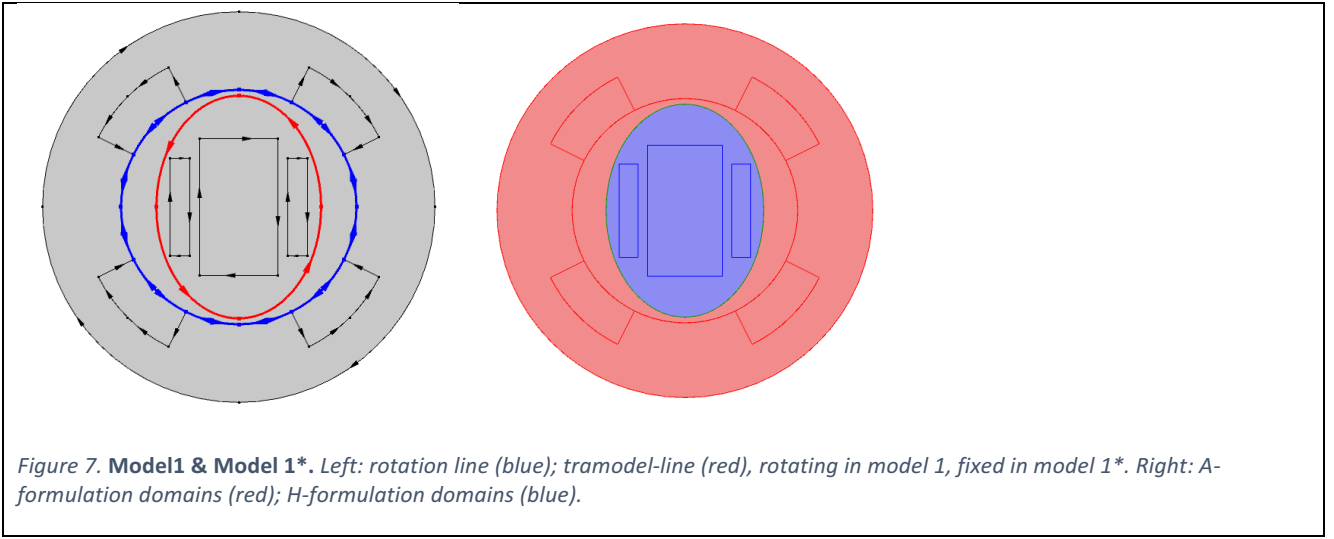

*Figure 7.* **Model1 & Model 1\*.** *Left: rotation line (blue); tramodel-line (red), rotating in model 1, fixed in model 1\*. Right: A-formulation domains (red); H-formulation domains (blue).*

Inside the rotating part we trace an arbitrary line (the red ellipse in the figure) separating the domains modelled with the *H*-formulations (inside) from those modeled with the *A*-formulation (outside). The rotation-line is therefore completely inside the *A*-formulation part and it is automatically a line of continuity of the magnetic potential $A_z$. In this configuration, the tramodel-line is rotating and *internal* to the rotation-line.

**H-formulation part.** The equations for the *H*-formulation part need to be formulated in the rotating system $(X, Y)$, i.e. the current density is given by the rotating coordinates derivative

(22)    $J = \partial_X K - \partial_Y H$

where $(H, K)$ are accordingly the components in the rotating system and the power-law (i) is applicable. On the tramodel-line the tangential magnetic field is given by $(\boldsymbol{t}_m = T\boldsymbol{t}_s)$

(23)    $Ht_m = t_X \cdot H + t_Y \cdot K$

The total currents in the two sections $S_1$ and $S_2$ of the winding will be imposed by means of two Integral constraints as done in (11)

$$\int_{S_1} J \mathrm{d}s = I_a(t) \qquad , \qquad \int_{S_2} J \mathrm{d}s = -I_a(t).$$

**A-formulation part.** The partial differential equation for the potential $A_z$ must be formulated in the fixed frame, so that we will have quantities of spatial type

(24)    $B_x = \partial_y A_z \quad B_y = -\partial_x A_z \quad E_z = -\partial_t A_z \quad J_z = \sigma E_z.$

Using the transformation (21) on the tramodel-line we can define the electric field in the rotating system. On the external boundary of the stator we impose the condition of magnetic insulation $A_z = 0$.

**Coupling.** The joining conditions (15) and (16) described above must now be formulated in the rotating system (because the tramodel-line is rotating) and therefore we will have on the tramodel-line

(25)    *A*-formulation: $Ht_m \cdot \text{test}(A_z)$

(26)    *H*-formulation: $E \cdot \text{test}(Ht_m)$

Using interpolating functions (shape functions) of Lagrange type 2[nd] order for the *A*-formulation part and Curl type 2[nd] order for the *H*-formulation part, the solution does not present any trouble and is very quickly obtained.

Figure 8 shows the continuity of the field on the tramodel-line and the typical current density distribution in the coil sections due to a power-law.

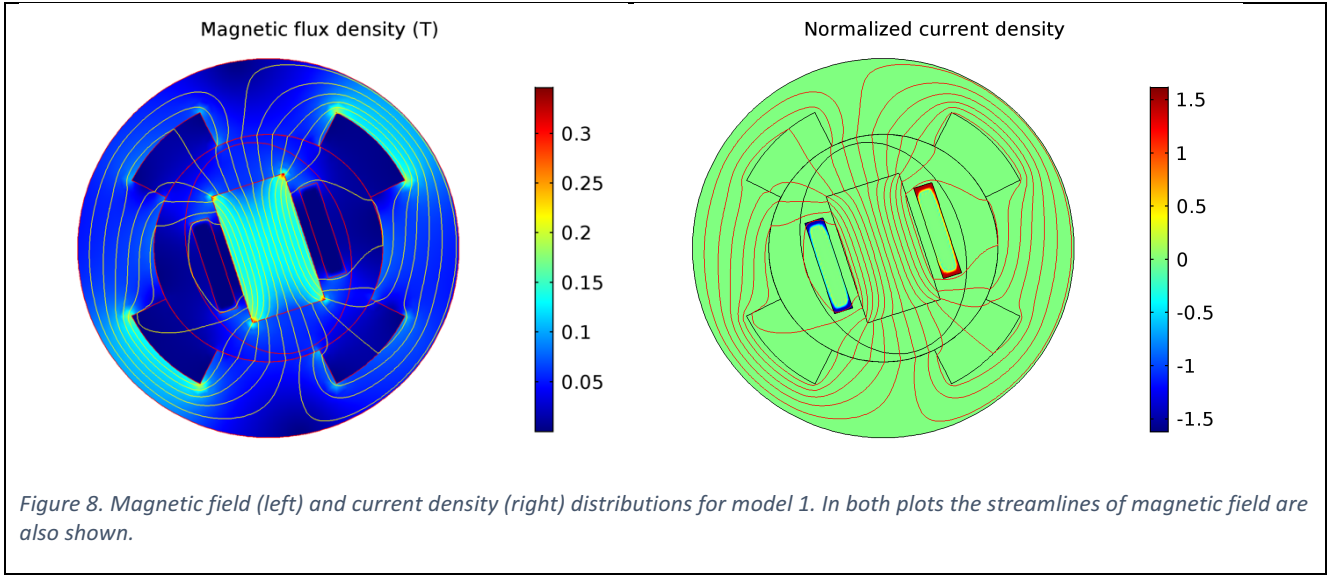

*Figure 8. Magnetic field (left) and current density (right) distributions for model 1. In both plots the streamlines of magnetic field are also shown.*

The power losses in the coils are simply given by the usual integral over the coil sections (*H*-formulation part)

$$P = \int_{S_1+S_2} \rho(J)J^2 \mathrm{d}s.$$

In order to verify the correctness of coupling (25) and (26) one has to plot the magnetic field components on the tramodel-line as computed by the two formulation as done in Figure 5.

## 3.3 Model 1* (fixed tramodel-line)

From a physical standpoint, nothing changes if, with the same geometry and current in the winding, one keeps the rotor fixed and makes the stator rotate in the opposite direction. In this case the tramodel-line is in the fixed part. Also fixed is the *H*-formulation part, which is spanned by the fixed frame. As a consequence, differently from the previous case, the equations will be

(27)     $J = \partial_x K - \partial_y H$

where now $(H, K)$ are the components in the fixed frame. On the tramodel-line the tangential magnetic field is given by

(28)     $Ht_s = t_x \cdot H + t_y \cdot K$

**Coupling.** Since the tramodel-line is now fixed, the joining conditions (24) and (25) must be recast in the fixed frame so we shall have

(29)     *A*-formulation: $Ht_s \cdot \text{test}(A_z)$

(30)     $H$-formulation: $E_z \cdot \text{test}(Ht_s)$

Since the electric field $E_z$ in the $A$-part is fixed too, it does not require the Lorentz correction $dE_z$ as in (20). The maps shown in Figure 9 are the same as those of Figure 8 , obviously rotated by the same angle in clockwise direction.

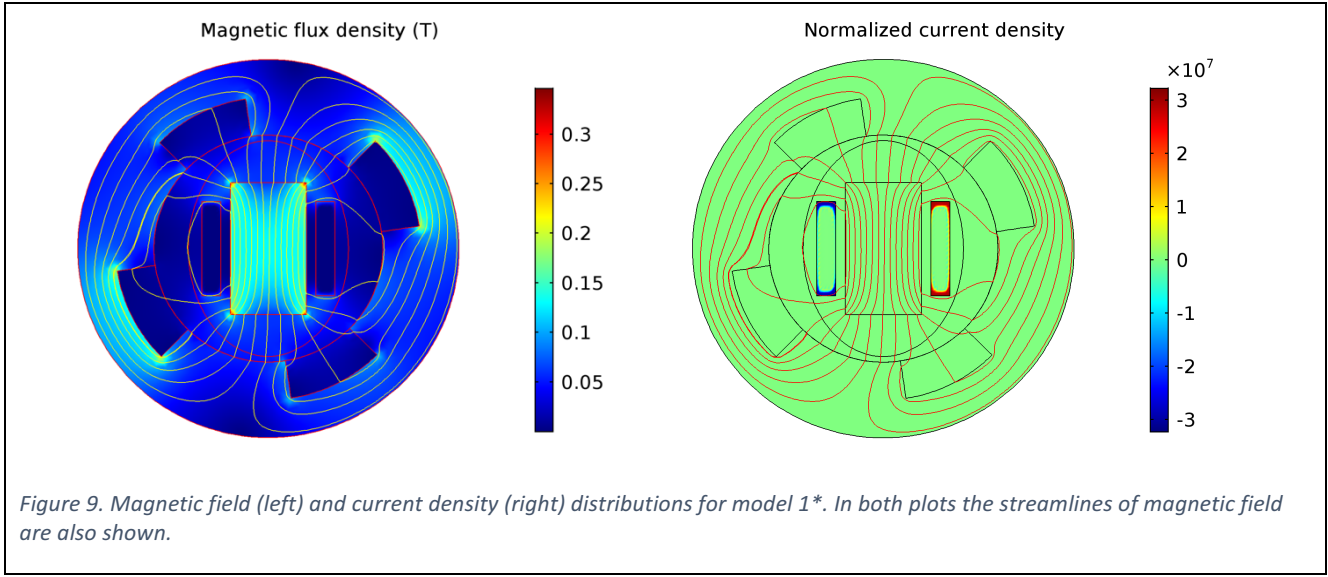

*Figure 9. Magnetic field (left) and current density (right) distributions for model 1\*. In both plots the streamlines of magnetic field are also shown.*

Again the plots of the magnetic field components given by both formulations on the tramodel-line can be used to test the correctness of the coupling conditions (29) and (30).

## 3.4 Model 2 (fixed tramodel-line)

Let us consider the case where the superconductors are in the external domain and the normal conductors in the internal one. Figure 10 shows the cross-section of a machine whose rotor is made of a normal conducting coil carrying a given current (DC or AC), whereas the fixed part consists of a coil made of two superconductors tapes located inside an iron tube.

Since the rotation-line has to be inside the $A$-formulation part, the tramodel-line consequently is external i.e. in the fixed part.

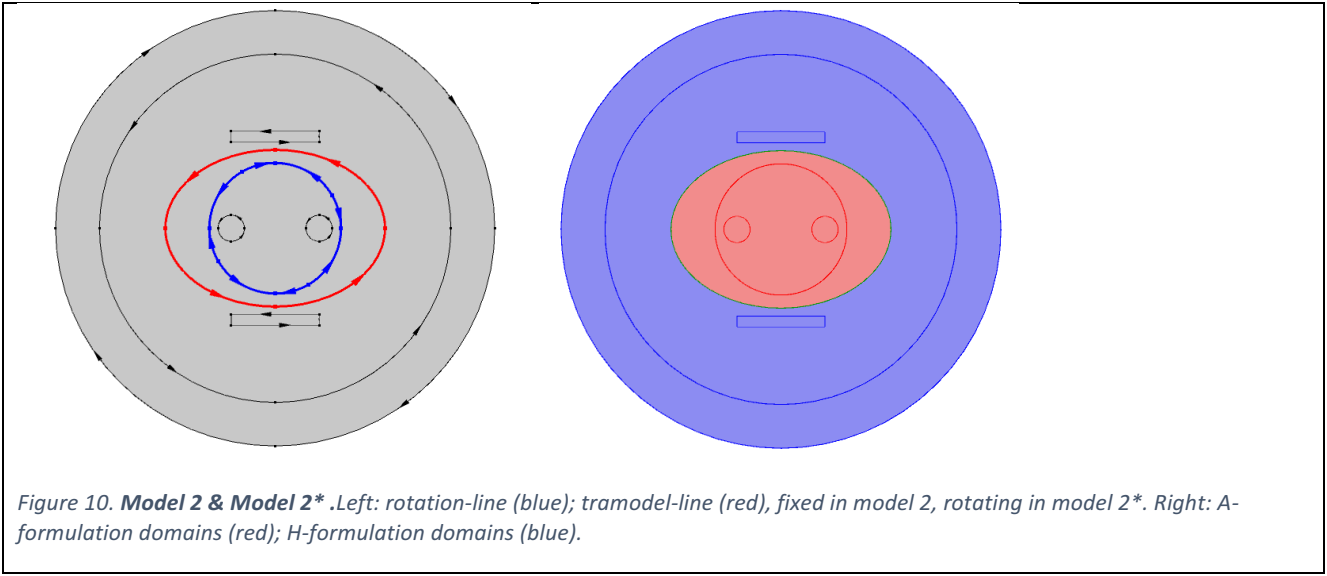

*Figure 10. **Model 2 & Model 2\*** .Left: rotation-line (blue); tramodel-line (red), fixed in model 2, rotating in model 2\*. Right: A-formulation domains (red); H-formulation domains (blue).*

**H-formulation part**. The equations have to be set in the fixed frame and the current density will be given by (27). On the tramodel-line we define the same tangential magnetic field of spatial type given by (28).

**A-formulation part**. Also the potential equation (1) has to be formulated in the fixed frame. The current in the rotating coil is imposed defining an external current density in the two sections

(31)     $J_e = \pm \sigma \, V_{coil}/L$

where $L$ is the longitudinal length (z-direction) of the coil; $V_{coil}$ is defined as in (12) by means of a global equation assuring the integral constraint

(32)     $I_{coil1,2} = \pm \int_{wire} (J_z + J_e) \mathrm{d}s = \pm I_a.$

**Coupling.** Since the tramodel-line is in the fixed part, the joining conditions are the same as in model 1\* with the swapping of the choice of the signs since the inversion of loop direction as seen by the two formulation (i.e. the swapping of the internal/external normal for the two part). So we have

(33)     A-formulation: $-Ht_s \cdot \text{test}(A_z)$

(34)     H-formulation: $-E_z \cdot \text{test}(Ht_s)$

Here also the electric field $E_z$ is of fixed type and does not requires the Lorentz term. Figure 11 clearly show the continuity of the field on the tramodel-line.

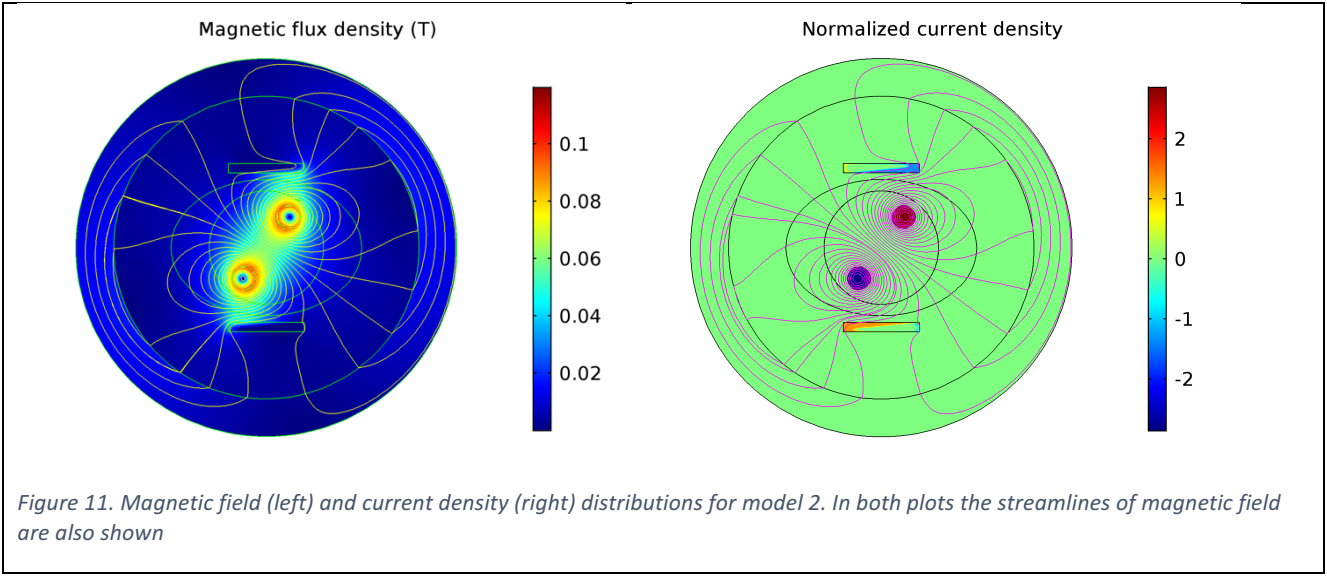

*Figure 11. Magnetic field (left) and current density (right) distributions for model 2. In both plots the streamlines of magnetic field are also shown*

The e.m.f. induced in the superconducting strips can be obtained as mean value of the electric field (per unit length) over their cross section

(35) $\quad V_{1,2} = \pm \frac{1}{L \cdot \text{section}} \int_{\text{section}} E \, ds$

Model 2 is specific for the case of a generator where the rotor field is generated by permanent magnets (easily modeled in the *A*-formulation part) and armature coils (fixed coils in the stator) made of superconductors.

## 3.5 Model 2* (rotating tramodel-line)

For completeness and to test the correctness of couplings we consider also the configuration obtained from the previous model exchanging the fixed part with the rotating part. The superconducting elements are in the portion of rotating part where the field is computed by the *H*-formulation, which is in the rotating frame. The rotor is fixed and here the magnetic field is formulated with the *A* potential in the fixed frame. The tramodel-line is entirely in the rotating part.

**Coupling**. Since the tramodel-line rotates, it is necessary to use joining conditions similar to those of model 1, but with the signs changed

(39) $\quad$ *A*-formulation: $-Ht_m \cdot \text{test}(A_z)$

(40) $\quad$ *H*-formulation: $-E_z \cdot \text{test}(Ht_m)$

where $Ht_m$ given by (23) and $E_Z$ given by (21) are the tangential magnetic field and the longitudinal electric field in the rotating reference system, respectively. As expected, the maps of Figure 12 are identical to the maps of Figure 11, apart from the rotation angle $\omega t$.

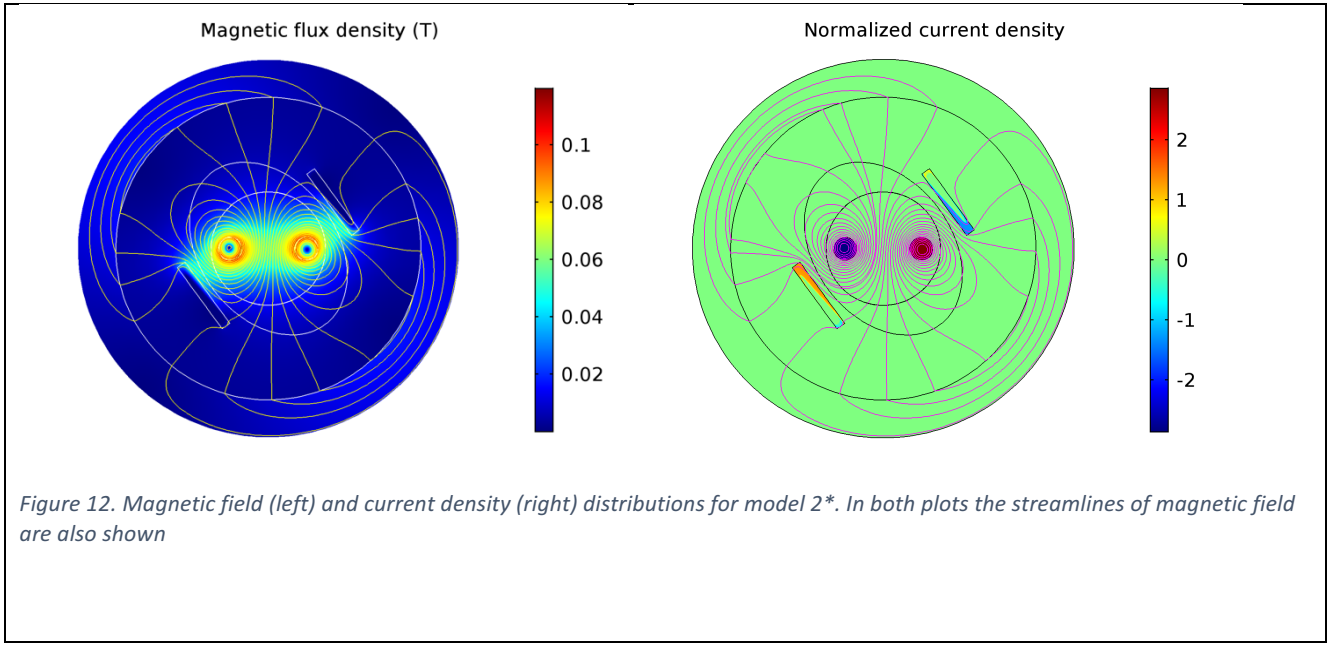

*Figure 12. Magnetic field (left) and current density (right) distributions for model 2*. In both plots the streamlines of magnetic field are also shown*

## 4 Conclusions

In this work, we developed a modeling framework for simulating the electromagnetic behavior of electrical machines with superconducting parts, whose material properties are described by a power-law resistivity. In this modeling framework:

1. The electromagnetic field of the region containing the superconductors are formulated with the partial differential equations for the components of the magnetic field – the well-known and widely used *H*-formulation using edge elements. The magnetic field in the remaining regions is formulated with the magnetic vector potential formulation with Lagrange elements.
2. The rotating boundary lies entirely in the *A*-formulation part.

The line separating the regions with different formulations – the tramodel-line – is the boundary were the conditions need to be "joined". A simple equality of the electromagnetic quantities formulated in the respective variables is not sufficient. The continuity of the tangential component of the magnetic field and of the perpendicular component of the electric field need to be forced by means of aptly-introduced Lagrange multipliers [12].

Depending on the simulated geometry and on the choice of the rotating and fixed parts, the tramodel-line can be in the rotating or in the fixed part, respectively. The joining conditions are therefore different in the two cases.

1. Rotating tramodel-line

    *A*-formulation: $\pm H t_m \cdot \text{test}(A_z)$

    *H*-formulation: $\pm E_z \cdot \text{test}(H t_m)$

where $Ht_m = t_X \cdot H + t_Y \cdot K$ and $E_Z = E_z - \omega\left(xB_x + yB_y\right)$ are the tangential component of the magnetic field and longitudinal component of the electric field in the rotating reference system (material), respectively.

2. Underline{Fixed tramodel-line}

$A$-formulation: $\pm Ht_s \cdot \text{test}(A_z)$

$H$-formulation: $\pm E_z \cdot \text{test}(Ht_s)$

where $Ht_s = t_x \cdot H + t_y \cdot K$ and $E_z = -\partial_t A_z$ are the tangential component of the magnetic field and longitudinal component of the electric field in the fixed reference system (spatial).

The signs to be used depend on the position of the $H$-formulation part (internal/external).

The shape and the position of the tramodel-line is largely arbitrary and some parts of the line may coincide with some elements of the geometry of the machine (for example, the terminals of polar expansions). Differently from the rotation-line, the tramodel-line does not need to be entirely in the air. This can be an advantage: a smart choice of the tramodel-line can avoid creating excessive mesh nodes in the region of rotation of the machine. However, one needs to check that all the arches composing the tramodel-line are equally oriented, so that the direction of the tangent is uniquely defined along the line. The $A$-formulation part is always situated in the spatial reference system, whereas the $H$-formulation part is situated in the material coordinate system if it rotates, or in the spatial system if it is fixed. The numerical solution of moving models does not present any additional difficulties with respect to the case of static models.

In all these models, some conductors are superconducting and other are normal and this fact determines the usage of the different formulations of magnetic field. In the case of all superconducting conductors (let say also the stator) a second tramodel line, in the fixed part, can be introduced in order to model with $H$-formulation also the fixed conductors. The fixed tramodel-line couplings (point 2 above) will be then applied on it. Only domains containing the rotation-line have to formulated by A-potential.

# Notes and References